\documentclass[apj]{emulateapj}
\usepackage{color}

\shorttitle{Fullerene/anthracene cluster cations}
\shortauthors{Zhen et al.}

\begin{document}
\title{Laboratory formation and photo-chemistry of fullerene/anthracene cluster cations}
\author{Junfeng Zhen$^{1,2,*}$, Weiwei Zhang$^{3,*}$, YuanYuan Yang$^{1,2,4}$, Qingfeng Zhu$^{1,2}$ Alexander G. G. M. Tielens$^{5}$} 

\affil{$^{1}$CAS Key Laboratory for Research in Galaxies and Cosmology, Department of Astronomy, University of Science and Technology of China, Hefei 230026, China} 
\affil{$^{2}$School of Astronomy and Space Science, University of Science and Technology of China, Hefei 230026, China}
\affil{$^{3}$Department of Mechanical and Nuclear Engineering, Pennsylvania State University, University Park, PA 16802, United States}
\affil{$^{4}$CAS Center for Excellence in Quantum Information and Quantum Physics, Hefei National Laboratory
	for Physical Sciences at the Microscale, and Department of Chemical Physics, University of Science and
	Technology of China, Hefei 230026, China}
\affil{$^{5}$Leiden Observatory, Leiden University, P. O. Box 9513, 2300 RA Leiden, The Netherlands} 

\email{jfzhen@ustc.edu.cn \& wzz2@psu.edu}

\begin{abstract}
    
Besides buckminsterfullerene (C$_{60}$), other fullerenes and their derivatives may also reside in space. In this work, we study the formation and photo-dissociation processes of astronomically relevant fullerene/anthracene (C$_{14}$H$_{10}$) cluster cations in the gas phase. Experiments are carried out using a quadrupole ion trap (QIT) in combination with time-of-flight (TOF) mass spectrometry. The results show that fullerene (C$_{60}$, and C$_{70}$)/anthracene (i.e., [(C$_{14}$H$_{10}$)$_n$C$_{60}$]$^+$ and [(C$_{14}$H$_{10}$)$_n$C$_{70}$]$^+$), fullerene (C$_{56}$ and C$_{58}$)/anthracene (i.e., [(C$_{14}$H$_{10}$)$_n$C$_{56}$]$^+$ and [(C$_{14}$H$_{10}$)$_n$C$_{58}$]$^+$) and fullerene (C$_{66}$ and C$_{68}$)/anthracene (i.e., [(C$_{14}$H$_{10}$)$_n$C$_{66}$]$^+$ and [(C$_{14}$H$_{10}$)$_n$C$_{68}$]$^+$) cluster cations, are formed in the gas phase through an ion-molecule reaction pathway. With irradiation, all the fullerene/anthracene cluster cations dissociate into mono$-$anthracene and fullerene species without dehydrogenation. The structure of newly formed fullerene/anthracene cluster cations and the bonding energy for these reaction pathways are investigated with quantum chemistry calculations. 

Our results provide a growth route towards large fullerene derivatives in a bottom-up process and insight in their photo-evolution behavior in the ISM, and clearly, when conditions are favorable, fullerene/PAH clusters can form efficiently. In addition, these clusters (from 80 to 154 atoms or $\sim$ 2 nm in size) offer a good model for understanding the physical-chemical processes involved in the formation and evolution of carbon dust grains in space, and provide candidates of interest for the DIBs that could motivate spectroscopic studies.

\end{abstract}

\keywords{astrochemistry --- methods: laboratory --- ultraviolet: ISM --- ISM: molecules --- molecular processes}

\section{Introduction}
\label{sec:intro}

Polycyclic aromatic hydrocarbon (PAH) molecules and their derivatives are believed to be very ubiquitous in the interstellar medium (ISM), where they are generally thought to be responsible for the strong mid-infrared (IR) features in the 3-17 $\mu$m range that dominate the spectra of most galactic and extragalactic sources in space \citep{all89, pug89, sel84, gen98}. The IR spectra of circumstellar and interstellar sources have also revealed the presence of buckminsterfullerene (C$_{60}$) in space \citep{cam10, sel10}, which is thought to be chemically linked to PAHs \citep{berne12,zhen2014}. In addition, several far-red diffuse interstellar bands (DIBs) are linked to the electronic transitions of C$_{60}$$^+$ \citep{cam15, wal15, cor17}. Using the Hubble Space Telescope, the recent study by \citet{cor19} confirms the presence of all three expected C$_{60}$$^+$ bands with strength ratios in agreement with those extrapolated from [C$_{60}$$-$He]$^+$ laboratory measurements. In addition, C$_{60}$ cations undergo Jahn-Teller distortion which removes the icosahedral symmetry of C$_{60}$. This eliminates various spectral symmetry-selection rules, and also leads to three different rotational constants \citep{sla03}, these two aspects have important implications for C$_{60}$$^+$/C$_{60}$ spectral observations in space \citep[and references therein]{lyk19}. Hence, understanding the formation and destruction processes of PAHs and fullerene, and their derivatives (e.g., fullerene/PAH clusters) has attracted much attention in the field of molecular astrophysics \citep{tie13, berne2015, omo16, gat16, can18}. 

Singular value decomposition analysis for the IR spectra of photo-dissociation regions (PDRs) has revealed the presence of a distinct emission component in the aromatic infrared bands (AIBs) that has been attributed to the presence of clusters of large molecules \citep{rapacioli05a, berne07}. Likewise, the extended red emissions (EREs), which dominates the visual spectra of reflection nebulae, has been attributed to luminescence by charged PAH clusters \citep{rhee07}. In addition, \citet{gar13a} proposed a possible relation between specific diffuse interstellar bands, including the strongest at 4428 \AA{}, and large fullerenes and buckyonions. \citet{gar13b} suggested fullerene/PAH adducts as candidates to the carriers of IR emission bands. And also, the formation and destruction of PAH clusters in PDRs has been studied by \citet{rapacioli06}. So far, the experimental evidence for the origin and evolution of clusters of large molecules in the gas phase has been lacking. 
	
In our previous studies, we have reported laboratory experiments on van der Waals bonded, PAH clusters and their photochemical evolution towards large PAH molecules in a bottom-up process \citep{zhen2018, zhang2019}. Given the presence of C$_{60}$ in PDRs such as NGC 7023 \citep{sel10}, studies of clusters involving C$_{60}$ have become of great interest as well. The processing of fullerene clusters by energetic ions \citep{gat16} is of relevance to interstellar shocks. Hence, in order to understand the formation and photochemical evolution of such species in PDRs, we will simulate and focus on the ultraviolet (UV) processing of clusters of PAHs and fullerene in the laboratory conditions. 

In addition, PAH/fullerene clusters offer a good approach to cosmic dust (very small grains) in terms of their scale size and their photochemical behavior \citep{cla99,omo16}. Indeed, the formation route for fullerenes and their derivatives provide a crucial anchor point to test models for the formation and evolution of carbon-rich dust, and recent experimental and quantum chemistry studies have started to elucidate this \citep{dun13, can19}.

It is known that fullerenes are electron-deficient poly-olefins that are able to form adducts with a number of different molecules \citep{kom99}. In particular, buckminsterfullerene (C$_{60}$) can react with catacondensed PAHs (e.g., acenes such as anthracene and pentacene) to form fullerene/PAH adducts via Diels$-$Alder cycloaddition reactions \citep{kom99,bri06,gar13a,gar13b,sat13}. Different yields of the neutral C$_{60}$/anthracene mono- and bis-adducts have been obtained in laboratory studies depending on the method employed in the production process, see e.g. \citet{gar13a,gar13b,cat14}. But the majority of these studies focused on the solid or liquid phase. Laboratory formation and photochemistry of (cationic) fullerene/PAH clusters in the gas phase have been barely investigated. Selective ion flow tube (SIFT) experiments reveal no reaction between C$_{60}$$^+$ and naphthalene but do show adduct formation with corannulene (C$_{20}$H$_{10}$) \citep{pet00}. \citet{dun13} also reported the formation of fullerene cluster cations resulting from the gas-phase interaction of C$_{60}$ and C$_{70}$ with coronene (C$_{24}$H$_{12}$) molecules under energetic conditions. In addition, laboratory studies of processing of van der Waals clusters of PAHs and of fullerenes in the gas phase by energetic ions (e.g., 24 keV O$^{2+}$ or 12 keV Ar$^{2+}$) have revealed the formation of chemically bonded large species through direct knock-out of carbon atoms \citep{zet10, zet13, del15}. 

In order to understand how (cationic) fullerenes aggregate with PAHs in the gas phase, we present an experimental and theoretical study on the photo-dissociation behavior of fullerene/anthracene cluster cations in the gas phase. The interaction of the fullerene cations, C$_{56}$$^+$, C$_{58}$$^+$, C$_{60}$$^+$, C$_{66}$$^+$, C$_{68}$$^+$ and C$_{70}$$^+$ with neutral anthracene are investigated. We select anthracene (C$_{14}$H$_{10}$, m/z=178) as an example of PAHs for this study, in view of its relatively high vapor pressure at room temperature. 

\begin{figure*}[t]
	\centering
	\includegraphics[width=\textwidth]{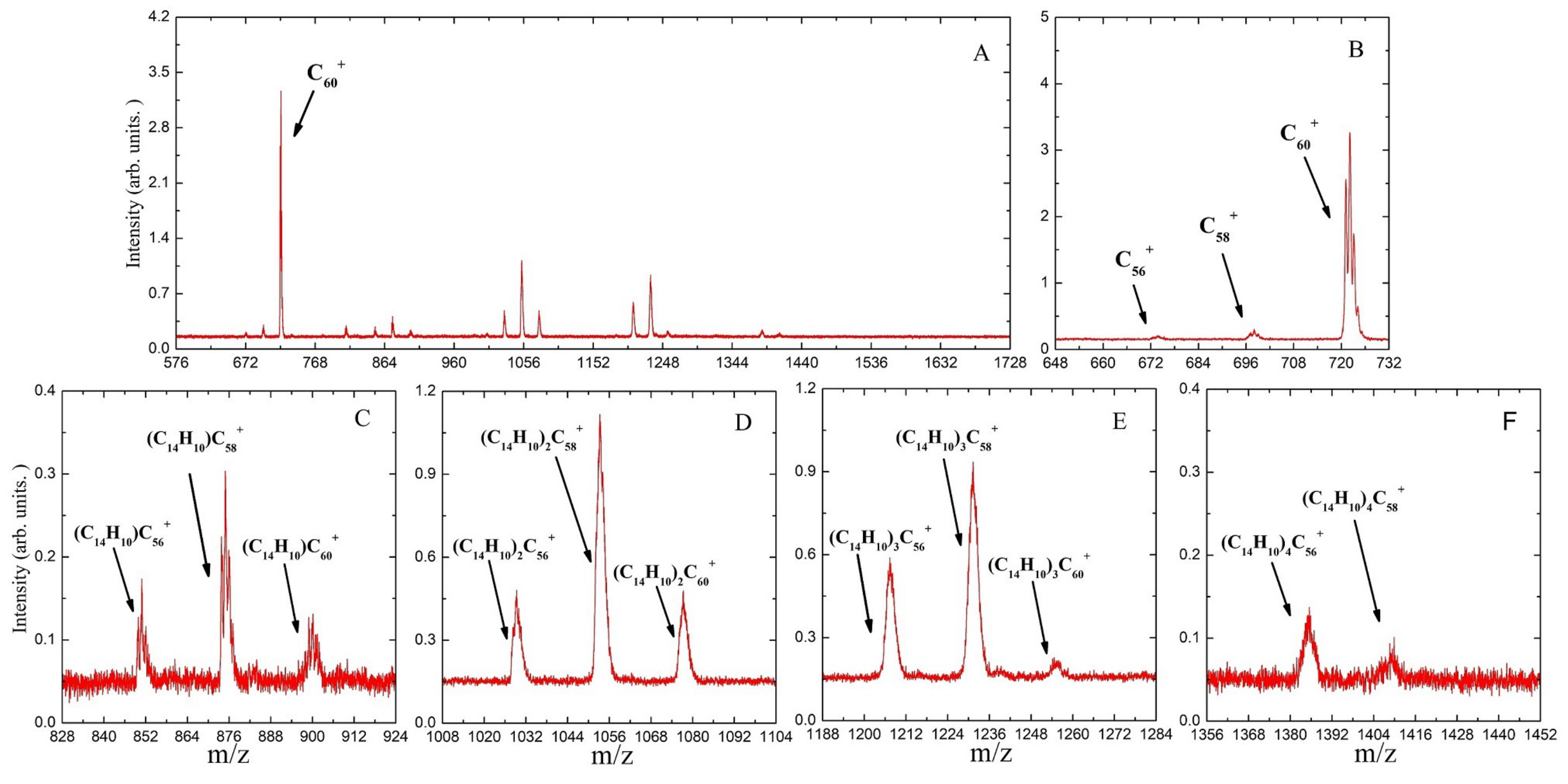}
	\caption{Upper panel (A): Mass spectrum of fullerene(C$_{56}$, C$_{58}$ and C$_{60}$)/anthracene cluster cations, without SWIFT and before laser irradiation; Panel (B-E): five zoom-in mass spectrum, revealing the presence of formed [C$_{56/58/60}$]$^+$, [(C$_{14}$H$_{10}$)C$_{56/58/60}$]$^+$, [(C$_{14}$H$_{10}$)$_2$C$_{56/58/60}$]$^+$, [(C$_{14}$H$_{10}$)$_3$C$_{56/58/60}$]$^+$ and [(C$_{14}$H$_{10}$)$_4$C$_{56/58}$]$^+$ cluster cations, respectively. 
	}
	\label{fig1}
\end{figure*}

\begin{figure}[t]
	\centering
	\includegraphics[width=\columnwidth]{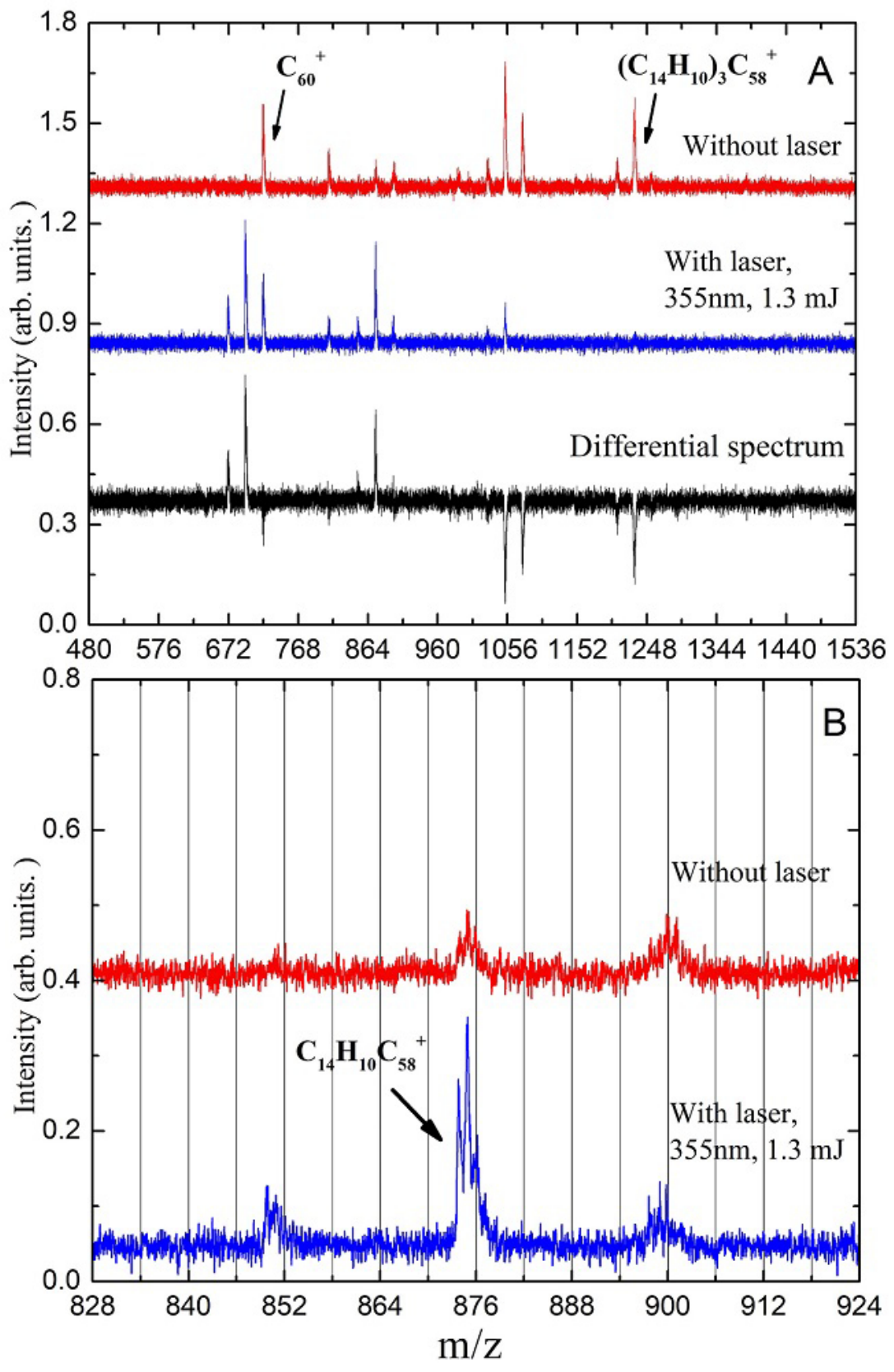}
	\caption{Upper Panel (A): Mass spectrum of fullerene(C$_{56}$, C$_{58}$ and C$_{60}$)/anthracene cluster cations trapped in QIT upon 355 nm irradiation at 1.3 mJ laser energy (irradiation times amounting to 0.5 s, from 4.4$-$4.9 s): without irradiation (red), with irradiation (blue), and the difference spectrum (black) of the irradiation and without irradiation experiments; Lower Panel (B): zoom in mass spectrum of [(C$_{14}$H$_{10}$)C$_{56/58/60}$]$^+$ cluster cations, without irradiation (red) and irradiated at 355 nm (blue) in the range of m/z=828-924;
	}
	\label{fig2}
\end{figure}

\begin{figure*}[t]
	\centering
	\includegraphics[width=\textwidth]{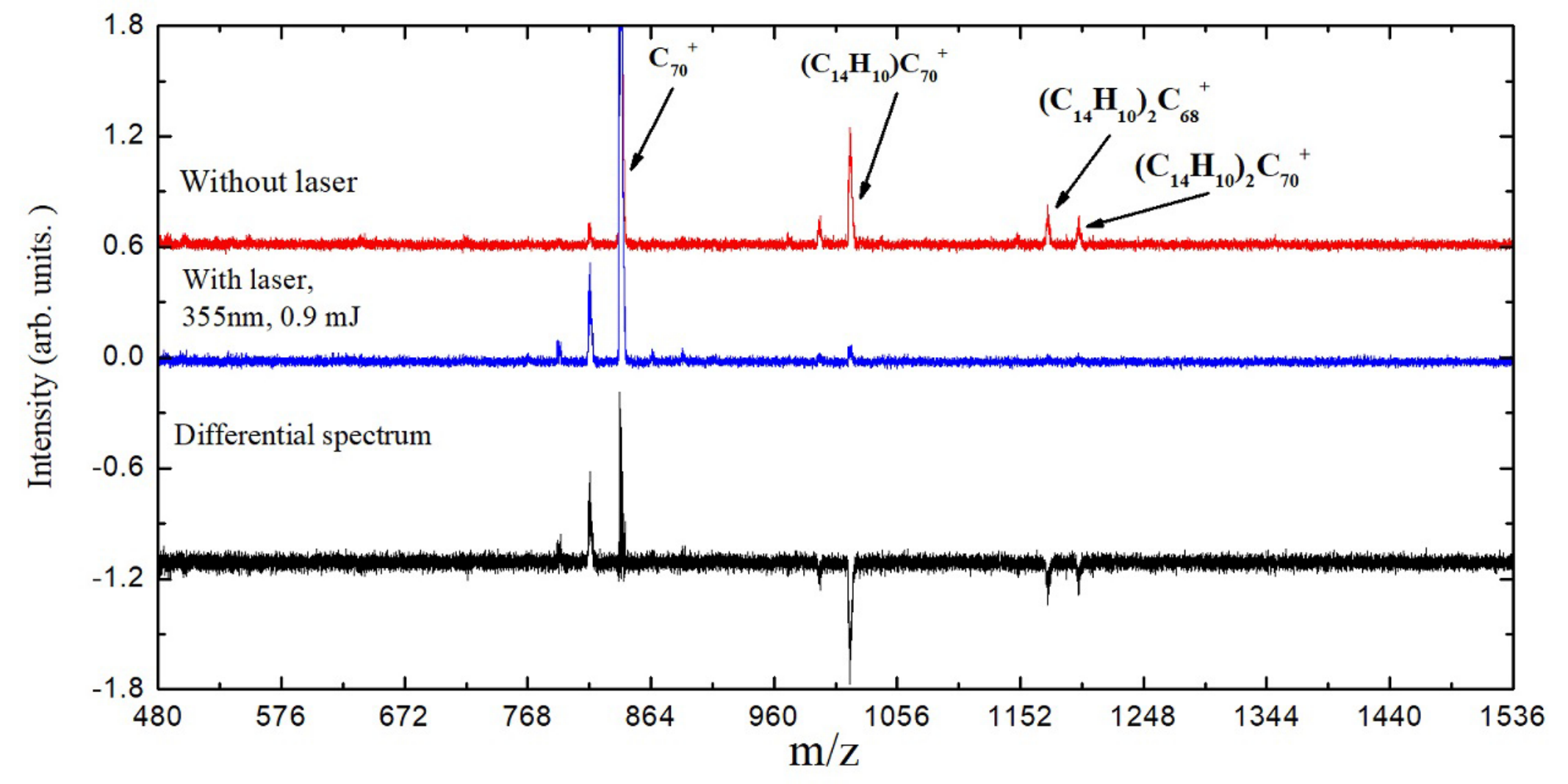}
	\caption{Mass spectrum of fullerene(C$_{66}$, C$_{68}$ and C$_{70}$)/anthracene cluster cations, without SWIFT and without irradiation (red), irradiated at 355 nm (blue) and the differential spectrum (black). In the no irradiation mass spectrum, revealing the presence of [(C$_{14}$H$_{10}$)C$_{66/68/70}$]$^+$ and [(C$_{14}$H$_{10}$)$_2$C$_{66/68/70}$]$^+$ clusters, respectively.
	}
	\label{fig3}
\end{figure*}

\begin{figure}[t]
	\centering
	\includegraphics[width=\columnwidth]{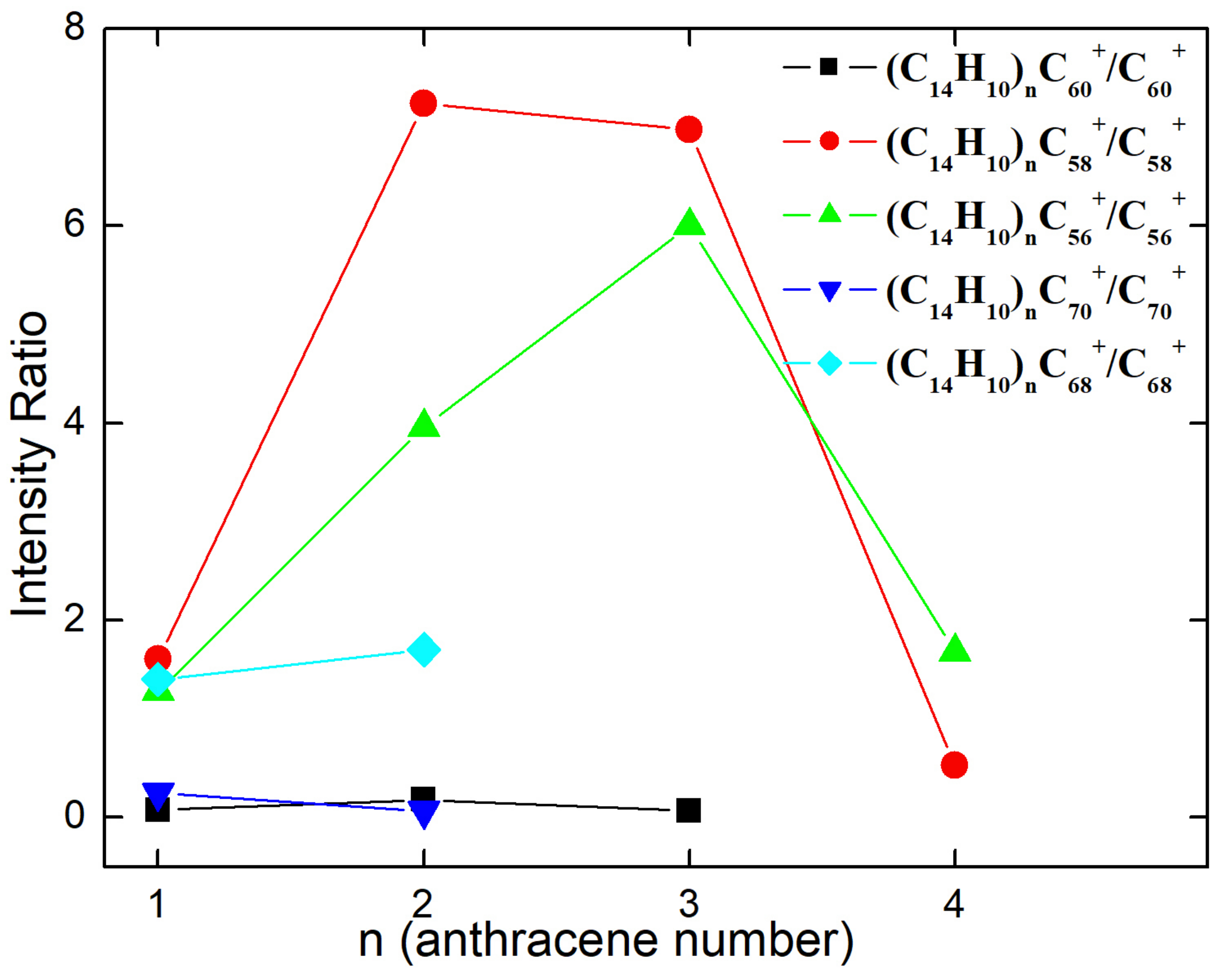}
	\caption{The intensity ratio of formed fullerene/anthracene cluster cations to fullerene cations: [(C$_{14}$H$_{10}$)$_n$C$_{60}$]$^+$/C$_{60}$$^+$, n=1, 2, 3; [(C$_{14}$H$_{10}$)$_n$C$_{58}$]$^+$/C$_{58}$$^+$ and [(C$_{14}$H$_{10}$)$_n$C$_{56}$]$^+$/C$_{56}$$^+$, n=1, 2, 3, 4; [(C$_{14}$H$_{10}$)$_n$C$_{70}$]$^+$/C$_{70}$$^+$ and [(C$_{14}$H$_{10}$)$_n$C$_{68}$]$^+$/C$_{68}$$^+$, n=1, 2. 
	}
	\label{fig4}
\end{figure}

\section{Experimental Methods}
\label{sec:exp}

Here, only a brief description of the experiment is provided. More detailed information on the experimental procedures is available in \citet{zhen2019}. First of all, a fullerene (C$_{60}$ or C$_{70}$) is evaporated by heating the powder (J\&K Scientific, with purity better than 99 \%) in the first oven at a temperature of $\sim$ 613 K. Subsequently, evaporated C$_{60}$ or C$_{70}$ molecules are ionized using electron impact ionization ($\sim$ 82 eV) and transported into the ion trap via an ion gate and a quadrupole mass filter. The high energy of the impacting electrons lead to fragmentation of the original fullerene through C$_2$ losses to form C$_{58}$$^+$ and C$_{56}$$^+$ or C$_{68}$$^+$ and C$_{66}$$^+$, respectively. 

A second oven (neutral molecules source, $\sim$ 300 K) is located under the trap to vaporize the molecules (anthracene power, J\&K Scientific, with a purity better than 99 \%), which can effuse continuously toward the center of the ion trap. In the ion trap, fullerene/anthracene cluster cations are formed by reaction between fullerene cations and neutral anthracene molecules. During this procedure, helium gas is introduced continuously into the trap via a leak valve to thermalize the ion cloud through collisions ($\sim$ 300 K). Adduct formation presumably occurs under our experimental operating conditions. The third harmonic of an Nd:YAG laser (INDI, Spectra-Physics), 355 nm, $\sim$ 6 ns, operated at 10 Hz, is used to irradiate the trapped, new formed, cluster cations. A beam shutter (Uniblitz, XRS$-$4) acts as a physical shield inside in the chamber and determines the interaction time of the light with the trapped ion clusters. The shutter is externally triggered to guarantee that the ion cloud is irradiated only for a specified amount of time during each cycle. A high precision delay generator (SRS DG535) controls the full timing sequence. 

Our setup operates with a typical frequency of 0.2 Hz, i.e., one full measuring cycle lasts 5.0 s. At the leading edge of the master trigger, the ion gate is opened (0.0-4.0 s), allowing the ion trap to fill for a certain amount of ions. During this time, the trapped ion reacts with anthracene molecules to form new cluster cations. Once these clusters are formed, the stored waveform inverse Fourier transform excitation (SWIFT) pulse is applied to isolate species within a given mass/charge (m/z) range (4.0-4.2 s) \citep{dor96}. Afterwards ($\sim$ 0.2 s), the beam shutter is opened to irradiate the ion cloud (4.4-4.9 s). At the end of irradiation, a negative square pulse is applied to the end cap of ion trap to accelerate the ions moving out of the trap and diffusing into the field–free TOF region, where the mass fragments can be measured. The mass spectrum results are shown in Figures 1-3. 

\section{Experimental results and discussion}
\label{sec:results}

The mass spectrum of the fullerene (C$_{56}$, C$_{58}$ and C$_{60}$)/anthracene cluster cations, without SWIFT isolation and before laser irradiation, is shown in Figure 1(A). Clearly, a series of peaks of fullerene/anthracene cluster cations are observed. As shown in Figure 1(B), except the main fullerene (C$_{60}$) mass peak (C$_{60}$$^+$, m/z=720), the mass spectra before irradiation reveal a small amount of residual fullerene (C$_{60}$) fragments (C$_{56}$$^+$, m/z=672 and C$_{58}$$^+$, m/z=696), due to the electron impact ionization and fragmentation \citep{zhen2014}. We note that the peak intensity of m/z=721 ($^{13}$C$^{12}$C$_{59}$$^+$) is stronger than m/z=720 ($^{12}$C$_{60}$$^+$) in here, as shown in Figure 1(B), which is off the natural carbon element abundance, i.e., $^{13}$C contained species have a stronger peak intensity than pure $^{12}$C species in here, the possible reason may due to the experimental setup conditions (quadrupole ion trap). 

We label the formed fullerene (C$_{56}$, C$_{58}$ and C$_{60}$)/anthracene cluster cations in four zoom-in mass spectra in the lower panel of Figure 1(C-F). The new formed cluster cations are shown as follows: In Figure 1(C), [(C$_{14}$H$_{10}$)C$_{56}$]$^+$ (m/z=850), [(C$_{14}$H$_{10}$)C$_{58}$]$^+$ (m/z=874) and [(C$_{14}$H$_{10}$)C$_{60}$]$^+$ (m/z=898); In Figure 1(D), [(C$_{14}$H$_{10}$)$_2$C$_{56}$]$^+$ (m/z=1028), [(C$_{14}$H$_{10}$)$_2$C$_{58}$]$^+$ (m/z=1052) and [(C$_{14}$H$_{10}$)$_2$C$_{60}$]$^+$ (m/z=1076); In Figure 1(E), [(C$_{14}$H$_{10}$)$_3$C$_{56}$]$^+$ (m/z=1206), [(C$_{14}$H$_{10}$)$_3$C$_{58}$]$^+$ (m/z=1230) and [(C$_{14}$H$_{10}$)$_3$C$_{60}$]$^+$ (m/z=1254); In Figure 1(F), [(C$_{14}$H$_{10}$)$_4$C$_{56}$]$^+$ (m/z=1384) and [(C$_{14}$H$_{10}$)$_4$C$_{58}$]$^+$ (m/z=1408). We note that, as observed above in the mass spectrum of C$_{60}$$^+$, the mass peak intensity pattern of the [(C$_{14}$H$_{10}$)C$_{58}$]$^+$ isotopologues does not reflect the natural abundance of carbon isotopes. Although the strongest peak is expected for ions that comprise only $^{12}$C (m/z = 874), a system issue of the experimental setup gives this characteristic to the peak caused by ions that contain a $^{13}$C atom. We consider that this issue does not affect the interpretation of our measurements because we do not observe any clear difference in the behavior of the isotopologues with regard to formation and dissociation. In addition, we also observe one \textquotedblleft extra" peak (m/z=810). While no assignments can be provided, we suspect that this peak might be formed as a side-product, due to contaminations in the ion trap chamber.

Importantly, the adducts of C$_{58}$$^+$ (e.g., [(C$_{14}$H$_{10}$)$_n$C$_{58}$]$^+$) always dominate the corresponding adducts of C$_{56}$$^+$ (e.g., [(C$_{14}$H$_{10}$)$_n$C$_{56}$]$^+$) and the latter is always at least as strong or dominate the corresponding adducts of C$_{60}$$^+$ (e.g., [(C$_{14}$H$_{10}$)$_n$C$_{60}$]$^+$). C$_{56}$ adducts only dominate over C$_{60}$ adducts for n$>$2. We will examine the formation behavior in the view of theoretical chemistry calculations in the next section.

As to the formation pathway of fullerene (C$_{56}$, C$_{58}$ and C$_{60}$)/anthracene cluster cations, we believe that the fullerene-derived cluster cations are formed by ion-molecule reaction pathways, i.e., C$_{56/58/60}$ cations $+$ neutral anthracene molecule. The reaction process between fullerene cations and neutral anthracene molecular occurs through sequential steps and add repeatedly anthracene groups to the surface of fullerene cages. Based on the obtained results, we propose the formation pathways as shown below:
\begin{small}
    \begin{eqnarray}
      &[\rm C_{56/58/60}]^+ &\stackrel{\rm C_{14}H_{10}}{\longrightarrow} [\rm {C_{14}H_{10}C_{56/58/60}}]^+\\
    &[\rm {C_{14}H_{10}C_{56/58/60}}]^+ &\stackrel{\rm C_{14}H_{10}}{\longrightarrow} [\rm ({C_{14}H_{10}){_2}C_{56/58/60}}]^+ \\
    &[\rm ({C_{14}H_{10}){_2}C_{56/58/60}}]^+ &\stackrel{\rm C_{14}H_{10}}{\longrightarrow} [\rm ({C_{14}H_{10}){_3}C_{56/58/60}}]^+ \\
    &[\rm ({C_{14}H_{10}){_3}C_{56/58}}]^+ &\stackrel{\rm C_{14}H_{10}}{\longrightarrow} [\rm ({C_{14}H_{10}){_4}C_{56/58}}]^+
    \end{eqnarray}
\end{small} 

Figure 2 (A) shows the resulting mass spectrum of trapped fullerene (C$_{56}$, C$_{58}$ and C$_{60}$)/anthracene cluster cations upon 355 nm irradiation at 1.3 mJ laser energies (irradiation times amounting to 0.5 s; i.e., typically $\sim$ 5 pulses). As we can see, the intensity of lower mass peaks increases while the higher mass peaks decrease under the laser irradiation (Figure 2, middle blue spectrum). For clarity, further detailed comparison are presented as a differential spectrum (lowest trace in Figure 2). In the differential mass spectrum in Figure 2(A), the intensity of C$_{60}$$^+$ is close to zero. The possible reason is that under laser irradiation, no C$_{60}$$^+$ product formed, because the ionization energy of C$_{60}$ is greater than that of anthracene. As such, the charge of cluster will localize on the anthracene molecular group, rather than on the C$_{60}$ molecular group and dissociation of the cluster will not lead to C$_{60}$$^+$. We will discuss this photo-dissociation behavior further with theoretical calculations in the next section.

We present a zoom in mass spectrum of [(C$_{14}$H$_{10}$)C$_{56/58/60}$]$^+$ cluster cations in Figure 2(B), with and without laser irradiation. We can see that only mono-anthracene molecular group dissociation products are formed, and there is no evidence for other fragmentation channels (e.g., dehydrogenation). Hence, we conclude that larger fullerene-PAH cluster cations shrink to smaller clusters by sequentially shedding anthracene molecules without dehydrogenation pathway. Accordingly, we propose the following photo-dissociation pathway for fullerene/anthracene cluster cations (the photo-dissociation pathway for [(C$_{14}$H$_{10}$)C$_{60}$]$^+$ will discuss later in the discussion section): 
\begin{small}
    \begin{eqnarray}
    &[\rm ({C_{14}H_{10}){_4}C_{56/58}}]^+ &\stackrel{\rm h\nu}{\rightarrow} \rm C_{14}H_{10} + \rm [({C_{14}H_{10}){_3}C_{56/58}}]^+\\
    &[\rm ({C_{14}H_{10}){_3}C_{56/58/60}}]^+ &\stackrel{\rm h\nu}{\rightarrow} \rm C_{14}H_{10} + \rm [({C_{14}H_{10}){_2}C_{56/58/60}}]^+\\
    &[\rm ({C_{14}H_{10}){_2}C_{56/58/60}}]^+ &\stackrel{\rm h\nu}{\rightarrow} \rm C_{14}H_{10} + \rm [{C_{14}H_{10}C_{56/58/60}}]^+\\
    &[\rm {C_{14}H_{10}C_{56/58}}]^+ &\stackrel{\rm h\nu}{\rightarrow} \rm C_{14}H_{10} + \rm [C_{56/58}]^+
    \end{eqnarray}
\end{small}  
For the fullerene (C$_{70}$) family, the typical mass spectrum of the fullerene (C$_{66}$, C$_{68}$ and C$_{70}$)/anthracene cluster cations are shown in Figure 3. Similar to Figure 1, without laser irradiation (Figure 3, upper red spectrum), besides C$_{66}$$^+$ (m/z=792), C$_{68}$$^+$ (m/z=816) and C$_{70}$$^+$ (m/z=840), we detect newly formed cluster cations labelled as: [(C$_{14}$H$_{10}$)C$_{66}$]$^+$ (m/z=970), [(C$_{14}$H$_{10}$)C$_{68}$]$^+$ (m/z=994) and [(C$_{14}$H$_{10}$)C$_{70}$]$^+$ (m/z=1018); [(C$_{14}$H$_{10}$)$_2$C$_{66}$]$^+$ (m/z=1148), [(C$_{14}$H$_{10}$)$_2$C$_{68}$]$^+$ (m/z=1172) and [(C$_{14}$H$_{10}$)$_2$C$_{70}$]$^+$ (m/z=1196). Based on the observed new species, we propose the formation reactions as:
\begin{small}
    \begin{eqnarray}
    &[\rm C_{66/68/70}]^+ &\stackrel{\rm C_{14}H_{10}}{\longrightarrow} [\rm {C_{14}H_{10}C_{66/68/70}}]^+\\
    &[\rm {C_{14}H_{10}C_{66/68/70}}]^+ &\stackrel{\rm C_{14}H_{10}}{\longrightarrow} [\rm ({C_{14}H_{10}){_2}C_{66/68/70}}]^+
    \end{eqnarray}
\end{small} 
The mass spectrum after irradiation is shown in Figure 3 (middle blue spectrum, 0.9 mJ). Further details on the photo-dissociation behavior of fullerene(C$_{66}$, C$_{68}$ and C$_{70}$)/anthracene cluster cations are presented as a differential spectrum (lower trace in Figure 3). Again, we do not observe the dehydrogenation process of these fullerene (C$_{66}$, C$_{68}$ and C$_{70}$)/anthracene cluster cations. Rather, larger clusters shrink to smaller clusters through mono$-$anthracene loss:
\begin{small}
    \begin{eqnarray}
    &[\rm ({C_{14}H_{10}){_2}C_{66/68/70}}]^+ &\stackrel{\rm h\nu}{\longrightarrow} \rm C_{14}H_{10} + [\rm {C_{14}H_{10}C_{66/68/70}}]^+\\
    &[\rm {C_{14}H_{10}C_{66/68/70}}]^+ &\stackrel{\rm h\nu}{\longrightarrow} \rm C_{14}H_{10} + \rm [C_{66/68/70}]^+
    \end{eqnarray}
\end{small}  
In order to compare the reactivity of C$_{58}$$^+$ and C$_{56}$$^+$ relative to C$_{60}$$^+$ in the cluster formation pathway, the intensity ratio of the fullerene/anthracene cluster cations to their parent fullerene cations are plotted in Figure 4: [(C$_{14}$H$_{10}$)$_n$C$_{60}$]$^+$/C$_{60}$$^+$, n=1, 2, 3; [(C$_{14}$H$_{10}$)$_n$C$_{58}$]$^+$/C$_{58}$$^+$ and [(C$_{14}$H$_{10}$)$_n$C$_{56}$]$^+$/C$_{56}$$^+$, n=1, 2, 3, 4; respectively. From the intensity ratio comparison, we conclude that C$_{58}$$^+$ and C$_{56}$$^+$ are more reactive towards adduct formation than C$_{60}$$^+$, and this is true for n=1, 2, 3. In addition, we note that, C$_{56}$$^+$ becomes more reactive than C$_{58}$$^+$ for n greater or equal to 4. Our conclusion is in line with previous studies of reactions of the fullerene cation (C$_{60}$$^+$) with cyclopentadiene \citep{bec97}. In these latter studies, C$_{58}$$^+$ and C$_{56}$$^+$ were shown to be more reactive to adduct formation than C$_{60}$$^+$. Nevertheless, we stress that, despite this low reactivity of C$_{60}$$^+$, we do see cluster formation of anthracene with C$_{60}$$^+$ up to n=3. For the C$_{70}$$^+$ family, the intensity ratio of [(C$_{14}$H$_{10}$)$_n$C$_{70}$]$^+$/C$_{70}$$^+$ and [(C$_{14}$H$_{10}$)$_n$C$_{68}$]$^+$/C$_{68}$$^+$, n=1, 2, are plotted in Figure 4. The C$_{70}$$^+$ family has a comparable adduct behavior to the C$_{60}$$^+$ family, and we conclude that C$_{68}$$^+$ is more reactive towards adduct formation than C$_{70}$$^+$. We will discuss the adduct behavior with theoretical chemistry calculations in the next section.

\section{Theoretical chemistry calculation results}
\label{sec:theoretical}

The theoretical calculations are carried out at the B3LYP \citep{bec92, lee88} level with the 6-31G(d, p) basis set, which is implemented in the Gaussian 16 program \citep{fri16}. To account for the weak interaction (i.e. van der Waals force) between fullerene and anthracene molecules, the dispersion-correction (D3) \citep{gri11} is included in this work. We mention here that the results reported here do not include the basis set-superposition error (BSSE) correction, which usually results in slightly reduced bond energies \citep{bas19}. In addition, we only carried out theoretical calculations for the fullerene (C$_{56}$, C$_{58}$ and C$_{60}$)/anthracene cluster cations system, due to the similarity in behavior for the fullerene (C$_{70}$) family. 

For the fullerene (C$_{56}$ and C$_{58}$) cations, we assume there is no carbon skeleton rearrangement (except for the C$_2$ loss at a local position) during the electron impact ionization and fragmentation process. After C$_2$ loss, there are two main isomers of C$_{58}$$^+$, namely 7 C-ring and 8 C-ring conformations, as shown in Figure 5. In agreement with earlier studies \citep{lee04, chen08, can19}, we found that the 7 C-ring isomer structure is more stable. Therefore, we only focus on this isomer in our following calculations. Based upon the C$_{58}$$^+$ study, after a further C$_2$ loss, we identify two isomers for C$_{56}$$^+$ similarly (Figure 5). And the double 7 C-rings conformation (the 7 C-rings are in opposite cage position) is more stable, which is therefore selected in the C$_{56}$$^+$ study.

To understand the details of formation process of the fullerene/anthracene cluster cations, we take C$_{60}$$^+$ $+$ anthracene, C$_{58}$$^+$ (7 C-ring) $+$ anthracene and C$_{58}$$^+$ (6 C-ring) $+$ anthracene as typical examples, to theoretically study the adduct reaction process. We follow the minimum energy pathway from the van der Waals cluster to the covalently bonded cluster, and at each step calculated the energy and the optimized structures. The energy and the optimized structure for the reactant, transition states (TS1 and TS2), intermediary, product for the reaction pathway between C$_{60}$$^+$ and anthracene, C$_{58}$$^+$ (7 C-ring) and anthracene and C$_{58}$$^+$ (6 C-ring) and anthracene are shown in Figure 6 and 7, and Table 1. 
	 
As shown in Figure 6, with B3LYP+D3 functional method, in the beginning, C$_{60}$$^+$ and anthracene form a van der Waals molecular complex (Initial, the exothermic energy is around -1.19 eV), and then it goes to an intermediary (Inter, -0.82 eV) through the first transition states (TS1, -0.85 eV) that pass the first activate barrier (0.37 eV). After that, the product (Product, the exothermic energy is around -1.31 eV) is formed through the second transition states (TS2, -0.73 eV) that pass the second activate barrier (0.08 eV). In the anthracene \textquotedblleft landing" on the C$_{60}$$^+$ process, the two carbon atoms from C$_{60}$ are puckered out of the cage surface, the structure of C$_{14}$H$_{10}$ is modified to allow the 9, 10 C-atoms to bond to the C-atoms from the fullerene \citep{sat13}. 

In order to check the accuracy of, e.g., the van der Waals interaction, M06-2X level are also calculated and the results are presented in Figure 6. In M06-2X functional calculation, the adduct process is similar to B3LYP+D3 functional method, and the calculation result are (intial, -1.25 eV), (TS1, -0.87 eV), (inter, -0.91 eV), (TS2, -0.78 eV) and (Product, -1.55 eV). The relative energy values obtained with M06-2X functional method all slightly increase as compared to the B3LYP+D3 functional method, especially for the covalently bonded (-1.55 eV to -1.31 eV) species. In agreement with \citep{sat13}, B3LYP+D3 performs almost as good as the M06-2X functional. For computational reasons, the B3LYP+D3 functional method was employed in this work for other fullerene/PAH cluster cations system.
	
For the interaction of C$_{58}$$^+$ with anthracene, due to the structure of C$_{58}$$^+$, there are two reaction pathways: one is \textquotedblleft landing" on the \textquotedblleft6 C-ring" and the other is \textquotedblleft landing" on the \textquotedblleft7 C-ring". The result of the calculation are presented in Figure 7(A) and (B), respectively. As shown in Figure 7(A), in the beginning, C$_{58}$$^+$ (7 C-ring) and anthracene form a van der Waals molecular complex (Initial, the exothermic energy is around -1.12 eV), and then it goes to an intermediary (Inter, -1.39 eV) through the first transition states (TS1, -1.07 eV) that pass the first activate barrier (0.05 eV). After that, the product (Product, the exothermic energy is around -1.31 eV) is formed through the second transition states (TS2, -1.01 eV) that pass the second activate barrier (0.38 eV). For the other possible route in the interaction of C$_{58}$$^+$ with anthracene, as shown in Figure 7(B), in the beginning, C$_{58}$$^+$ (6 C-ring) and anthracene form a van der Waals molecular complex (Initial, the exothermic energy is around -0.83 eV), and then it goes to an intermediary (Inter, -0.34 eV) though the first transition states (TS1, -0.31 eV) that pass the first activate barrier (0.52 eV). After that, the product (Product, the exothermic energy is around -1.05 eV) is formed though the second transition states (TS2, -0.14 eV) that pass the second activate barrier (0.20 eV). 

For elucidate the difference in the chemical behavior of the three fullerene cations with anthracene, especially for the newly formed multi-anthracene adducted clusters, we present in Figure 8 the optimized structures of the covalently bonded clusters, [(C$_{14}$H$_{10}$)$_{(1-3)}$C$_{60}$]$^+$ (panel A), [(C$_{14}$H$_{10}$)$_{(1-4)}$C$_{58}$]$^+$ (panel B) and [(C$_{14}$H$_{10}$)$_{(1-4)}$C$_{56}$]$^+$ (panel C). To the covalently bonded clusters, mono-anthracene and fullerenes that are connected by two C-C single bonds in which the two (blue) C-atoms are from C$_{60}$ and two (red) C-atoms are from anthracene. All four of these C-atoms are in sp$^3$ hybridization where for the two C-atoms from anthracene one of the sp$^3$ bonds is a C-H bond. As expected \citep{bri06} and the discussion presented above, the more reactive 9 \& 10 C-atoms of anthracene are involved in the covalent bond formation. 

In Figure 8(A), [C$_{14}$H$_{10}$C$_{60}$]$^+$ and the optimized structure of [(C$_{14}$H$_{10}$)$_2$C$_{60}$]$^+$ and [(C$_{14}$H$_{10}$)$_3$C$_{60}$]$^+$ are obtained. For these structures, additional anthracene is also added to (normal) 6 C-rings. Clearly, all the reactions are exothermic, with -1.3, -1.3 and -1.0 eV, respectively. 

Similar to C$_{60}$/anthracene, [C$_{14}$H$_{10}$C$_{58}$]$^+$ and the optimized structure of [(C$_{14}$H$_{10}$)$_2$C$_{58}$]$^+$, [(C$_{14}$H$_{10}$)$_3$C$_{58}$]$^+$ and [(C$_{14}$H$_{10}$)$_4$C$_{58}$]$^+$ (shown in Figure 8(B)) are obtained. Base on the obtained result in Figure 7, for these multi anthracene adducted clusters $-$ and analogously to C$_{60}$/anthracene and C$_{58}$/anthracene $-$ additional anthracene is added to (normal) 6 C-rings. Clearly, all the reactions are also exothermic, with -1.3, -1.2, -1.1 and -0.8 eV, respectively. 

As shown in Figure 8(C), similar to C$_{58}$/anthracene cluster cations, the structure of [C$_{14}$H$_{10}$C$_{56}$]$^+$ consists of one mono-anthracene molecule and one C$_{56}^+$,  connected by two C-C single bonds. Again, two carbon atoms from the fullerene are in sp$^3$ hybridization from one of the 7 C-rings, and two C-atoms from anthracene are in sp$^3$ hybridization with an additional C-H bond. Accordingly, the optimized structures of [(C$_{14}$H$_{10}$)$_2$C$_{56}$]$^+$, [(C$_{14}$H$_{10}$)$_3$C$_{56}$]$^+$ and [(C$_{14}$H$_{10}$)$_4$C$_{56}$]$^+$ are obtained and shown in Figure 8(C). For [(C$_{14}$H$_{10}$)$_2$C$_{56}$]$^+$, the second anthracene molecule is added on another 7 C-ring. For [(C$_{14}$H$_{10}$)$_3$C$_{56}$]$^+$ and [(C$_{14}$H$_{10}$)$_4$C$_{56}$]$^+$, anthracene molecules are added to normal 6 C-rings. Clearly, all the reactions are exothermic, with -1.7, -1.3, -0.9 and -0.8 eV, respectively. 

Comparison of the optimized structures of the covalently bonded clusters of the three fullerenes with anthracene reveals clear differences. Specifically, with three anthracene on C$_{60}$$^+$ cage surface, the spherical shape of C$_{60}$$^+$ is almost unchanged. When we introduce four anthracene on the C$_{56}$$^+$ cage surface, the shape of the C$_{56}$$^+$ cage shows significant modification changing from a spherical cage to a more tetrahedral cage. We surmise that this difference reflects the very rigid structure of C$_{60}$$^+$ as compared to much more pliable C$_{58}$$^+$ and C$_{56}$$^+$ cages. 

\section{Discussion}
\label{sec:dis}

The experiments show that the fullerene cations, C$_{58}$$^+$ and C$_{56}$$^+$, react much more readily with anthracene than C$_{60}$$^+$. Likewise, C$_{68}$$^+$ and C$_{66}$$^+$ react more readily than C$_{70}$$^+$. Following \citep{boh16}, we can attribute this to the enhanced curvature of the surfaces of C$_{58}$$^+$ \&\ C$_{56}$$^+$ and C$_{68}$$^+$ \&\ C$_{66}$$^+$ with respect to C$_{60}$$^+$ and C$_{70}$$^+$ respectively. In the theoretical calculations, we can separate the formation process of fullerene/anthracene cluster cations into two stage: the first stage is from fullerene cation + anthracene to the van der Waals cluster (Initial), the second stage is from the van der Waals cluster to covalent bonded cluster (Product). The formation of both the van der Waals cluster and the covalently bonded species are energetically downhill from the reactants. The binding energies of the van der Waals cluster and the covalently bonded species are very similar for both species, [C$_{14}$H$_{10}$C$_{60}$]$^+$ and [C$_{14}$H$_{10}$C$_{58}$]$^+$. However, there is a substantial energy barrier in the transition from the van der Waals cluster involving C$_{60}$$^+$ to the covalently bonded cluster, while there is hardly a barrier involving C$_{58}$$^+$. While, in either process, this barrier is submerged, we surmise that it does play an important role in the reaction process; That is, possibly the [C$_{14}$H$_{10}$C$_{60}$]$^+$ cluster is quickly trapped in the van der Waals, but with insufficient energy to overcome the energy barrier to the covalently bonded species. In contrast, after trapping in the van der Waals well, [C$_{14}$H$_{10}$C$_{58}$]$^+$ can still react to form a covalently bonded species. We then further surmise that the [C$_{14}$H$_{10}$C$_{60}$]$^+$ van der Waals cluster does not survive in the TOF mass spectrometer acceleration zone, while the covalently bonded [C$_{14}$H$_{10}$C$_{58}$]$^+$ species does. In this view, the small amount of [C$_{14}$H$_{10}$C$_{60}$]$^+$ represents van der Waals clusters that \textquotedblleft survived" in the acceleration process. 

We do note that, in addition, at low temperatures (around 10 K), the substantial dipoles of C$_{56}$$^+$ (1.1 Debye) and C$_{58}$$^+$ (1.26 Debye) can be expected to enhance the reaction rate coefficient of these species by a factor of 2 compared to C$_{60}$$^+$ (0 Debye). However, at 300 K in our experimental condition, the effect is negligible \citep{smi11}. Finally, it should be emphasized that reactions in the liquid and solid state can lead to covalently bonded structures \citep{gar13a, gar13b}.

As to the subsequent dissociation pathway initiated by the laser irradiation in Figure 2, the clusters evolve towards breaking the bond between the fullerene and the anthracene groups. From our previous studies, we have shown that the dissociation energy of H-loss is generally $\sim$ 2.0 eV for aliphatic carbon of PAHs cluster cations \citep{zhen2018}, which is larger than the calculated bond energy of anthracene and fullerene ($\sim$ from 0.8 to 1.7 eV as shown above). Hence, in agreement with the experiments, loss of anthracene molecule should dominate over H-loss.

In addition, as shown as below, the charge transfer can happen between the C$_{60}$ cation and anthracene (exothermic reaction pathway, + 0.37 eV, equation 13) through cluster cations. In contrast, the charge exchange reaction of C$_{58}$$^+$ with anthracene is endothermic by 0.27 eV which is thermodynamically unfavorable (equation 14).
\begin{scriptsize}
	\begin{eqnarray}
	&\rm C_{14}H_{10} + \rm [C_{60}]^+ &\rightarrow \rm [C_{14}H_{10}C_{60}]^+ \stackrel{\rm h\nu}{\rightarrow} \rm C_{14}H_{10}^+ + \rm C_{60} + \rm 0.37 eV\\
	&\rm C_{14}H_{10} + \rm [C_{58}]^+ &\rightarrow \rm [C_{14}H_{10}C_{58}]^+ \stackrel{\rm h\nu}{\rightarrow} \rm C_{14}H_{10}^+ + \rm C_{58} - \rm 0.27 eV
    \end{eqnarray}
\end{scriptsize} 

\begin{figure}[t]
	\centering
	\includegraphics[width=\columnwidth]{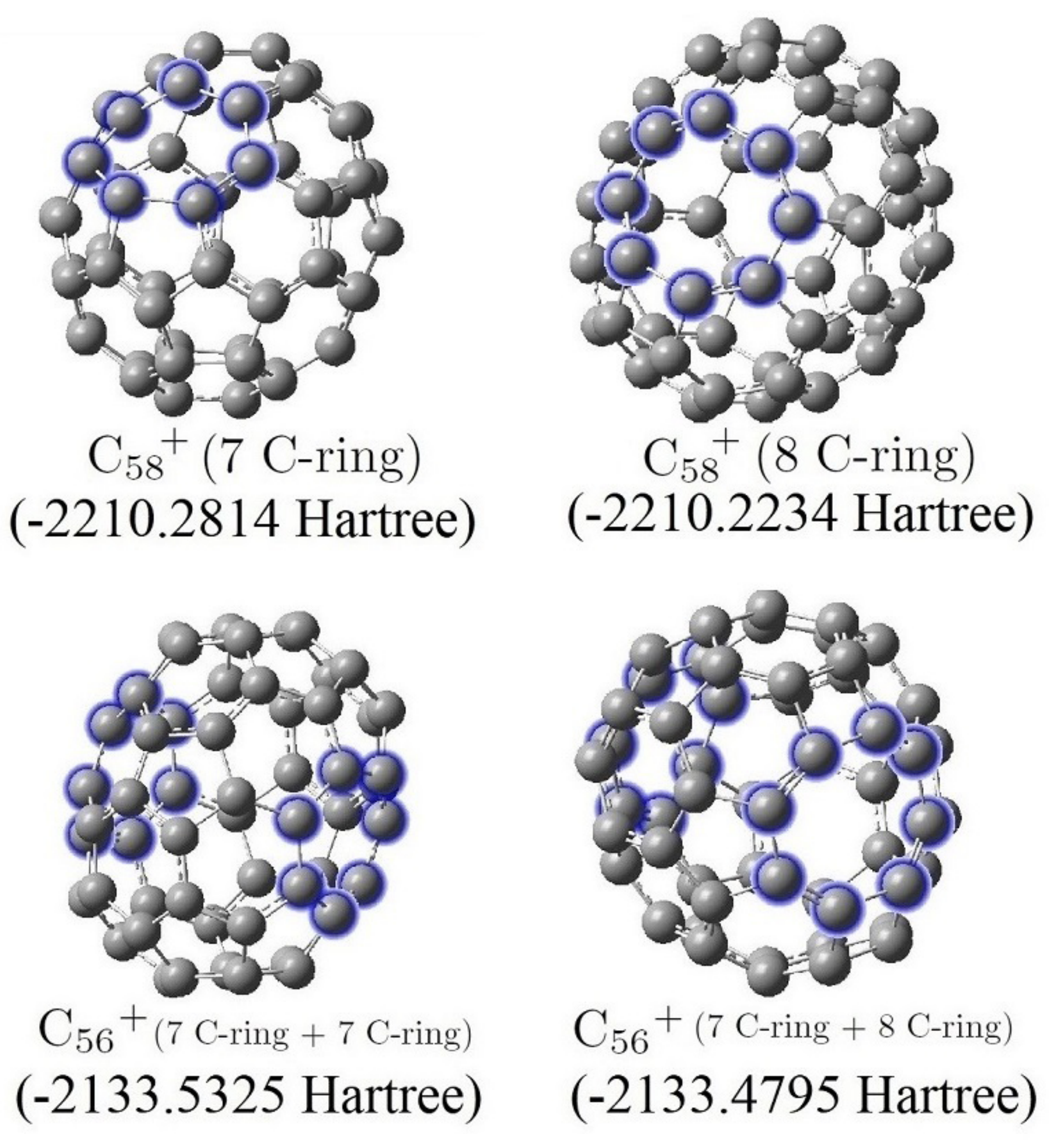}
	\caption{The optimized structure of C$_{56}$$^+$ and C$_{58}$$^+$, blue carbon for 7 C-ring or 8 C-ring.
	}
	\label{fig5}
\end{figure}

\begin{figure}[t]
	\centering
	\includegraphics[width=\columnwidth]{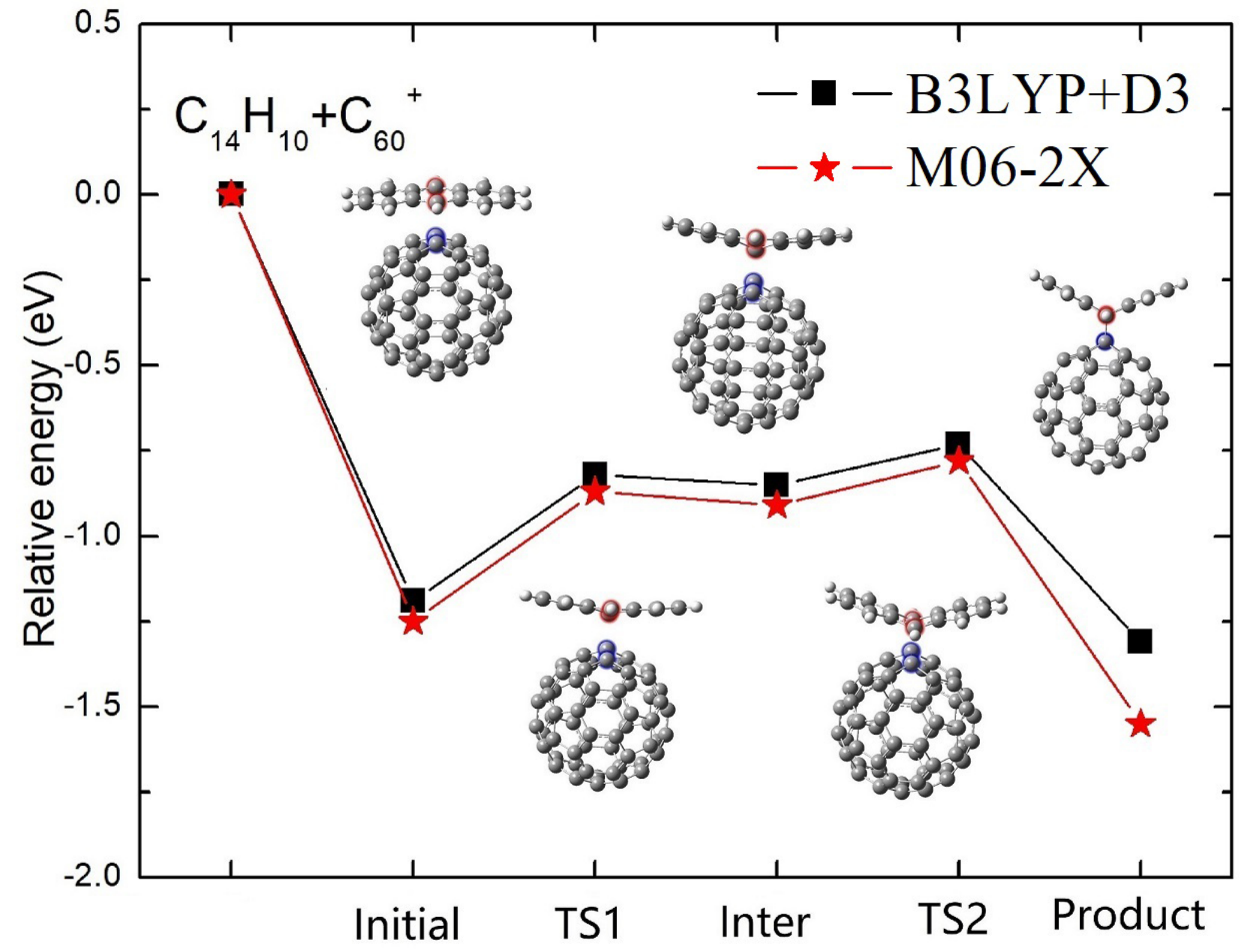}
	\caption{The reactant, transition states, intermediary, product, and the energy for the reaction pathway between C$_{60}$$^+$ and anthracene with calculation B3LYP+D3 and M06-2X functional method, respectively.
	}
	\label{fig6}
\end{figure}

\begin{figure}[t]
	\centering
	\includegraphics[width=\columnwidth]{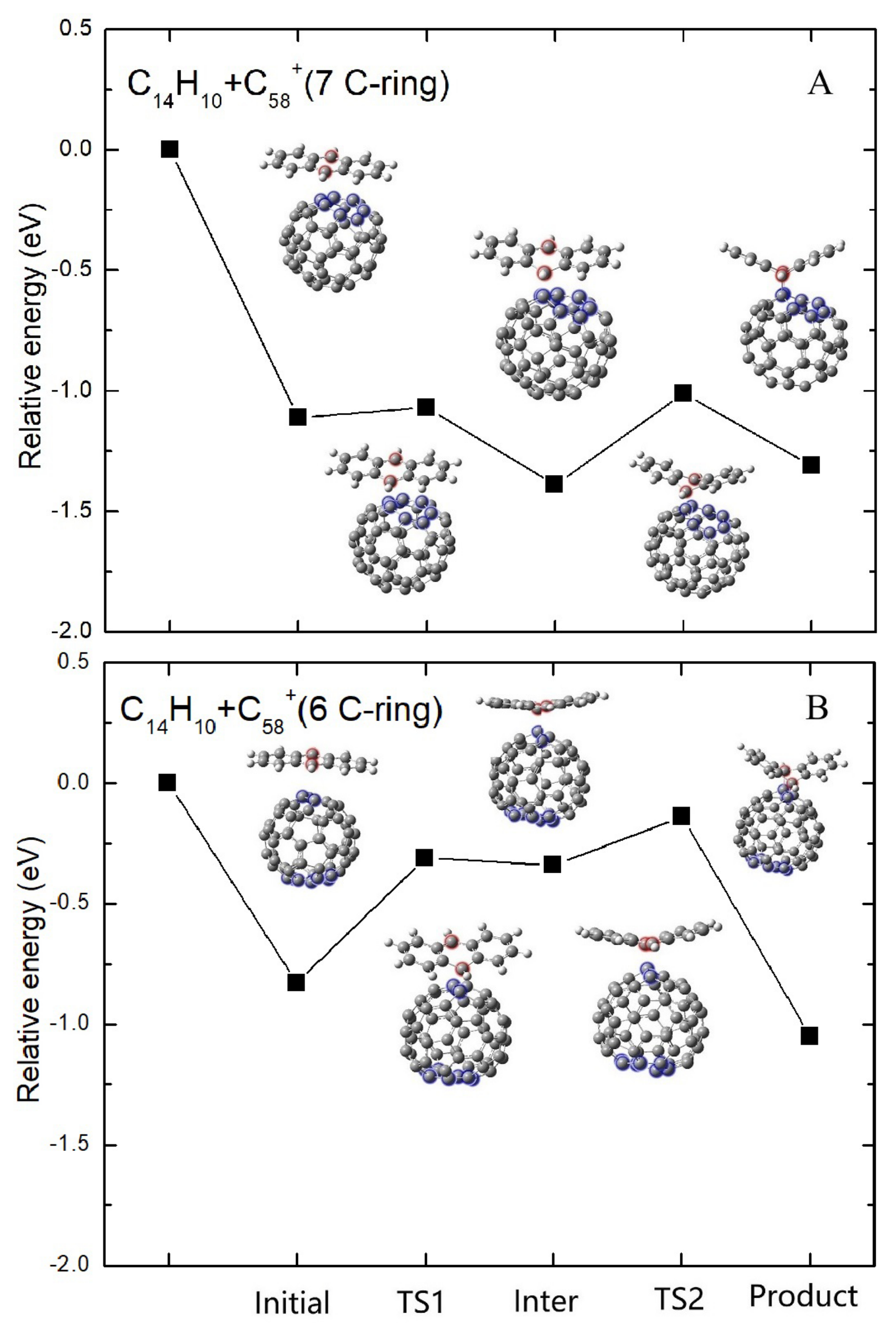}
	\caption{The reactant, transition states, intermediary, product, and the energy for the reaction pathway between C$_{58}$$^+$ (7 C-ring) and anthracene (panel A) and C$_{58}$$^+$ (6 C-ring) and anthracene (panel B), respectively, with B3LYP+D3 functional method.
	}
	\label{fig7}
\end{figure}

\begin{table*}
	\centering
	\caption{The energy for the reactant, transition states (TS1 and TS2), intermediary, product for the reaction pathway between C$_{60}$$^+$ and anthracene, C$_{58}$$^+$ (7 C-ring) and anthracene and C$_{58}$$^+$ (6 C-ring) and anthracene.
		\label{BenchCalcs}}
	\begin{tabular}{cccccccccccc}
		\hline
		\multicolumn{1}{c}{}&\multicolumn{4}{c}{C$_{60}$$^+$ and anthracene}&\multicolumn{2}{c}{C$_{58}$$^+$ (7 C-ring) and anthracene}&\multicolumn{2}{c}{C$_{58}$$^+$ (6 C-ring) and anthracene}\\
		\hline
		&\multicolumn{2}{c}{B3LYP+D3 level}&\multicolumn{2}{c}{M06-2X level}&\multicolumn{2}{c}{B3LYP+D3 level}&\multicolumn{2}{c}{B3LYP+D3 level}\\
		\hline
		&Hartree&eV&Hartree&eV&Hartree&eV&Hartree&eV\\
		\hline
		Reactant&	-2825.828330	&0.00&-2825.04131&	0.00	&-2749.492420&	0.00&	-2749.492420&	0.00\\
		Initial&	-2825.872179&	-1.19&-2825.08727&	-1.25&	-2749.533379&	-1.12&	-2749.522858&	-0.83\\
		TS1&	-2825.858382&	-0.82&-2825.07328&	-0.87&	-2749.531803&	-1.07&	-2749.503947&	-0.31\\
		Inter&	-2825.859637&	-0.85&-2825.07484&	-0.91&	-2749.543600&	-1.39	&-2749.504750&	-0.34\\
		TS2&	-2825.85503&	-0.73&-2825.07006&	-0.78&	-2749.529612&	-1.01&	-2749.497699&	-0.14\\
		Product&	-2824.876509&	-1.31&-2825.09826&	-1.55&	-2749.540684&	-1.31&	-2749.530882&	-1.05\\
		\hline
	\end{tabular}
	\\
\end{table*}

\begin{figure*}[t]
	\centering
	\includegraphics[width=\textwidth]{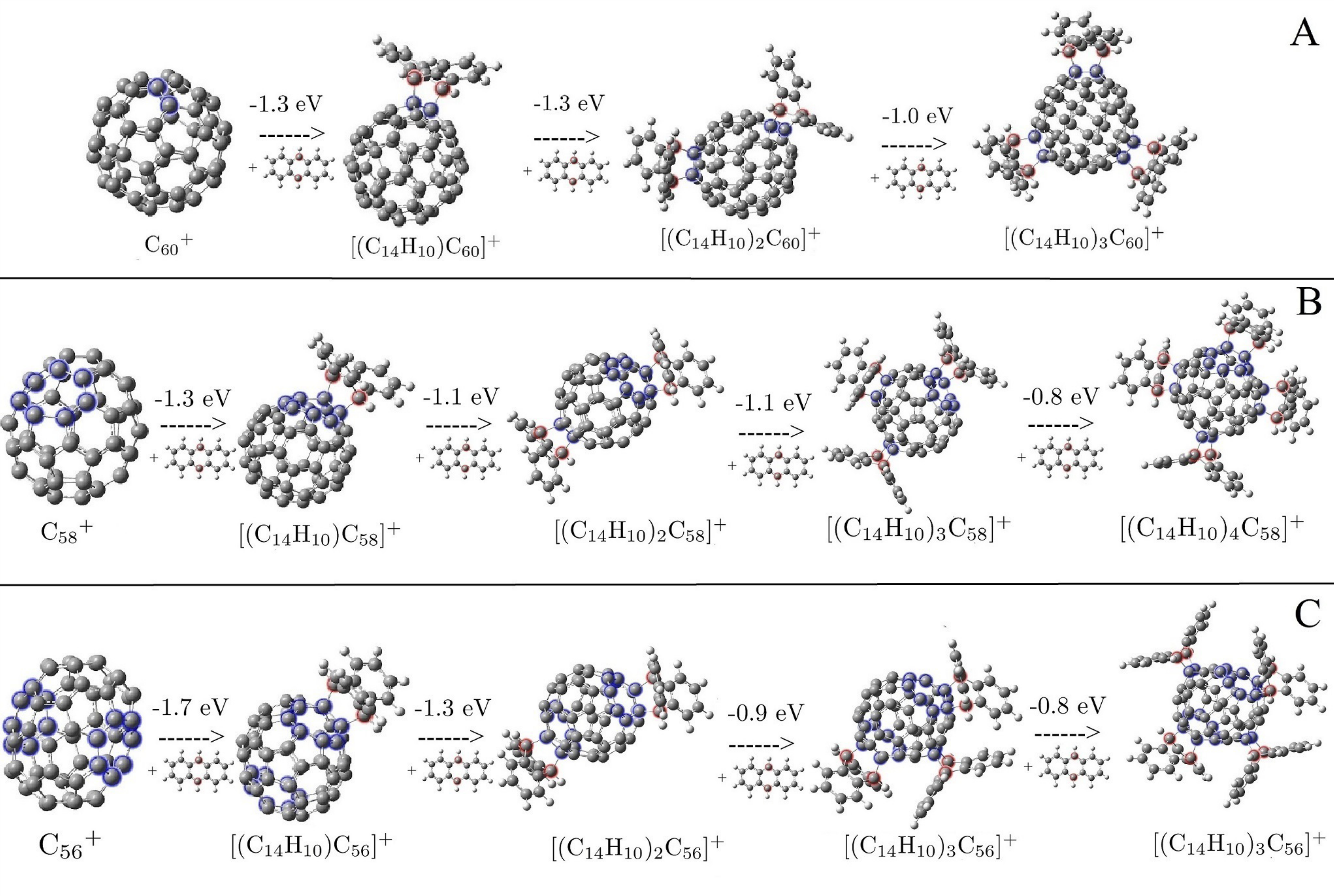}
	\caption{The formation reaction pathway for [(C$_{14}$H$_{10}$)$_n$C$_{60}$]$^+$, n=1, 2, 3 in panel (A); the formation reaction pathway for [(C$_{14}$H$_{10}$)$_n$C$_{58}$]$^+$, n=1, 2, 3, 4 in panel (B); the formation reaction pathway for [(C$_{14}$H$_{10}$)$_n$C$_{56}$]$^+$, n=1, 2, 3, 4 in panel (C). (Blue carbon is from fullerene group and red carbon is from C$_{14}$H$_{10}$ group for C$-$C bond)
	}
	\label{fig8}
\end{figure*}

\section{Astronomical implications}
\label{sec:discussion}

We experimentally and theoretically investigated the formation and photo-chemistry processes of a series of large fullerene derivatives (e.g., fullerene-PAH derived clusters). Gas-phase reactions between fullerene (e.g., C$_{56/58}$, C$_{66/68}$ and C$_{60/70}$ cations) and PAHs (e.g., anthracene) occur in our experimental setup, which provide new insights into the evolution of fullerene (bottom-up growth) in the radiation fields in the ISM. \citet{pet00} presented an experimental study of C$_{60}$$^+$ adduct reaction with one anthracene or corannulene (C$_{20}$H$_{10}$) molecule. In these experiments, adduct formation with anthracene did not occur but did occur with corannulene. Our experiments also indicate very inefficient adduct formation of anthracene with C$_{60}$$^+$ (compared to C$_{58}$$^+$ and C$_{56}$$^+$). \citet{dun13} demonstrated the cluster cations ([C$_{24}$H$_{10}$$-$C$_{60}$]$^+$ and [C$_{24}$H$_{10}$$-$C$_{70}$]$^+$) formation resulting from gas-phase interaction of C$_{60}$ and C$_{70}$ with coronene (C$_{24}$H$_{12}$) under energetic conditions. In our study, we build upon these studies by investigating the adduct formation behavior of C$_{56}$$^+$/C$_{58}$$^+$/C$_{60}$$^+$ and C$_{66}$$^+$/C$_{68}$$^+$/C$_{70}$$^+$ with anthracene, revealing the much greater reactivity of the smaller fullerenes derived from C$_{60}$ and C$_{70}$ by successive C$_2$ losses. In addition, for the first time, we obtained the multi-PAHs adducting on the fullerene surface (e.g., [(C$_{14}$H$_{10}$)$_4$C$_{58}$]$^+$, four anthracene molecules on the C$_{58}$ cage surface as one super large molecule clusters, with 154 atoms and $\sim$ 2 nm in size).

The much greater reactivity of C$_{58}$$^+$ and C$_{56}$$^+$ to adduct formation as compared to C$_{60}$$^+$ is in line with the study of \citet{bec97}, which reported experimental evidence for the heightened chemical reactivity of C$_{58}$$^+$/C$_{56}$$^+$ relative to C$_{60}$$^+$ in the Diels-Alder reaction with cyclopentadiene. This difference in behavior has been related to the more pliable cage structure of the smaller fullerenes as suggested by \citet{pet00,bec97} in their study of the Diels-Alder reaction of these fullerene cages with cyclopentadiene.

Our present study indicates that the smaller fullerene cations, C$_{58}$$^+$ and C$_{56}$$^+$ (C$_{68}$$^+$ and C$_{66}$$^+$), form adducts with PAHs much more readily than C$_{60}$$^+$ (C$_{70}$$^+$). Hence, if these smaller fullerene are present in space, formation of covalently bonding fullerene-based clusters could produce an extended family of large molecules (together with the van der Waals cluster of C$_{60}$$^+$ with anthracenes). Likewise, these types of clusters may play a role in the IR spectral complexity of circumstellar environments where C$_{60}$ has been shown to be prominent \citep{cam10, slo14, ots14}. \citep{ben12} could not explain satisfactorily the relative intensities of IR emission bands attributed to C$_{60}$ in PNes. They suggested that other substances, e.g., C$_{70}$, could contribute to some of the bands, thus causing the inconsistencies they observed. Because the spatial distributions of fullerenes and PAHs do not overlap in PNes, we do not propose fullerene/PAH adducts as such contributors. Nevertheless, species formed by reaction between fullerene cations and molecules found in PNes may be involved.

In addition, it has been suggested that PAH clusters play a role in the extended red emission prominent in many interstellar and circumstellar environments \citep{rhee07}. These fullerene/PAHs adducts formed in our experiments may be relevant for this emission as well. In addition, the covalent bond formation in the clusters considered here may be an important step in the formation of larger carbon grains \citep{dun13}.  

In this paper, we study the subsequent photo-chemically driven evolution of such fullerene/PAHs cluster cations. The calculated binding energy is $\sim$1.5 eV which is less than the binding energy of sp$^3$ (and sp$^2$) H-atoms in PAHs. In space, the weakest link is expected to go first and fragmentation after UV excitation will lead to loss of a PAH molecule from the cluster. Based upon this low binding energy and using the density of states of C$_{60}$ and PAHs as a guide, we estimate that absorption of a single 6 eV photon - which are readily available in PDRs - will be sufficient to lead to fragmentation \citep{tie05}. As this estimate scales with the internal energy per atom, larger clusters will require concomitantly more energetic photons. For much larger clusters, multi-photon events can then still lead to fragmentation in a PDRs environment \citep{berne2015}.

\section{Conclusions}
\label{sec:concl}

The first experimental results on the formation and photo-chemical process of large fullerene/anthracene cluster cations in the gas phase are presented, which reveal a general cluster formation process for fullerene/PAHs cluster cations, i.e., constructed a series of fullerene-PAH derived cluster molecules. The cluster formation process is especially true for the smaller fullerene (C$_{56}$, C$_{58}$ and C$_{66}$, C$_{68}$) cations. In agreement with earlier studies involving reactions with cyclopentadiene \citep{bec97, boh16}, we conclude that C$_{58}$$^+$ and C$_{56}$$^+$ are much more reactive towards cluster formation than C$_{60}$$^+$. Quantum chemistry calculations demonstrate that these newly formed cluster species can be quite stable (the binding energy $\sim$ 1.3 eV), which provides a possible further evolution route of fullerene complexation in the ISM. Subsequent photo-processing (355nm is used in here) can weed down those fullerene/anthracene cluster cations to their most stable forms and back to fullerene and mono-anthracene molecular groups again. 

\acknowledgments

This work is supported by the Fundamental Research Funds for the Central Universities and from the National Science Foundation of China (NSFC, Grant No. 11421303 and Grant No. 11590782). Studies of interstellar chemistry at Leiden Observatory are supported through a grant by the Netherlands Organisation for Scientific Research (NWO) as part of the Dutch Astrochemistry Network and through the Spinoza premie.

\end{document}